\documentclass[11pt]{iopart}
\usepackage[T1]{fontenc}
 \usepackage[latin9]{inputenc}
 \usepackage{verbatim}
 \usepackage{graphicx}
 \usepackage{amssymb}
\usepackage{color}

\begin{document}
\title{Quantum state transfer in a $XX$ chain with impurities}

\author{Analia Zwick, Omar Osenda}

\address{Facultad de Matem\'atica, Astronom\'ia y F\'isica, Universidad Nacional de C\'ordoba, and IFEG-CONICET, 
   Ciudad Universitaria, X5016LAE C\'ordoba, Argentina}

\ead{zwick@famaf.unc.edu.ar, osenda@famaf.unc.edu.ar}

\begin{abstract}
One spin excitation states are involved in the transmission of quantum
states and entanglement through a quantum spin chain, the localization
properties of these states are crucial to achieve the transfer of
information from one extreme of the chain to the other. We investigate
the bipartite entanglement and localization of the one excitation
states in a quantum $XX$ chain with one impurity. The bipartite entanglement
is obtained using the Concurrence and the localization is analyzed
using the inverse participation ratio. Changing the strength of the
exchange coupling of the impurity allows us to control the number
of localized or extended states. The analysis of the inverse participation
ratio allows us to identify scenarios where the transmission of quantum
states or entanglement can be achieved with a high degree of fidelity.
In particular we identify  
a regime where the transmission of quantum states between the extremes of the
chain is executed in a short transmission time $\sim N/2$, where $N$ is the
number
of spins in the chain, and with a large fidelity.
\end{abstract}



\section{Introduction }

Since the first works dealing with the entanglement shared by pairs
of spins on a quantum chain, the translational invariance of the chain
(and its states) has been exploited to facilitate the analysis of
the problem \cite{wootters2000,wootters2001,nachtergaele2006}. Anyway,
there is a number of problems which do not possess the property of
being translationally invariant: semi-infinite chains, chains with
impurities \cite{osenda2003} or, in a more abstract sense, random
quantum states \cite{zyczkowski2004}. These problems have localized
quantum states whose properties strongly differ from those of translationally
invariant quantum states.

Localized quantum states can be used to storage quantum information
\cite{apollaro2007} and play an important role in the propagation of entanglement through a quantum spin chain \cite{apollaro2006}.
This kind of states also appears in some models of quantum computers
in presence of static disorder \cite{Georgeot2000}.

Since the localization of a quantum state is a global property it
seems natural to study its properties using a global entanglement
measure as, for example, the one proposed by Meyer and Wallach \cite{mewa2002}. Giraud {\em et al.}~\cite{giraud2007} derived exact expressions
for the mean value of the Meyer-Wallach entanglement for localized
random vectors and studied the dependence of this measure with the
localization length of the states. Viola and Brown \cite{viola2007}
studied the relationship between generalized entanglement and the
delocalization of pure quantum states. Of course there are other
possibilities to study the relationship between localization of quantum
states and entanglement. The bipartite entanglement and localization
of one-particle states in the Harper model has been addressed by Li
{\em et al.}~\cite{li2004}, the entanglement entropy at localization
transitions is studied in \cite{jia2008} and the localized
entanglement in one-dimensional Anderson model in~\cite{li2004b}.

In many proposals of quantum computers the qubit energies can be often
individually controlled, this corresponds to controllable disorder
of a spin system. Besides, in these models, the effective spin-spin
interaction is usually strongly anisotropic, it varies from the Ising
coupling in nuclear magnetic resonance and other systems \cite{chuang1998}
to the $XY$-type or the $XXZ$-type coupling in some Josephson-junction-based
systems \cite{makhlin2001}. The localization properties of one and
two excitation states in the $XXZ$ spin chain with a defect was studied
with some detail by Santos and Dykman \cite{santos2003}, but they
did not study the entanglement of the one and two excitation states.

In this paper we are interested in the behaviour of the localization
and the bipartite entanglement of the pure eigenstates of a quantum
chain with one impurity located in one extreme. It is well know that
the presence of one impurity results in the presence of a localized
state. If the strength of the impurity is large enough the energy
of the localized state lies outside the band of magnons, also known
as one spin excitation states \cite{santos2003}. The one spin magnons
in a homogeneous chain are extended states \cite{santos2003}.

As we will show, if the localization of a given state is measured
with the inverse participation ratio there are two kinds of localized
states, a) exponentially localized states that lie outside the band
of magnons, and b) localized states that lie inside the band, whose
number depends on the length of the chain and the strength of the
impurity. This second kind plays a fundamental role in the transmission
of quantum states through the chain. In most quantum state transfer
protocols the state to be transferred is localized at one end of the
quantum chain and the transmission is successful when the time evolution
of the system produces an equally localized state at the other end
of the chain. So it seems natural to investigate the time evolution
of a localization measure to gain some insight about the problem of
quantum state transfer.

So, the analysis of  the time evolution 
of the inverse participation ratio,
when the initial state consists in a single excitation located in
one impurity, allows the identification of scenarios where the transmission
of quantum states can be achieved for (comparatively) short times
and with a very good fidelity. In this sense we extend some results
obtained by W\'ojcik {\em et al.}~\cite{wojcik2005}.

The paper is organized as follows, in Section II we present the $XX$
model describing the quantum spin chain with a impurity. In Section
III we analyze in some detail the spectrum of the one spin excitations
and the eigenstates. In Section IV we present the results obtained
for the inverse participation ratio for each one spin excitation eigenstate
while the bipartite entanglement of the eigenstates is analyzed in
Section V. Finally, in Section VI, we discuss the relationship between
localization and transmission of quantum states.

\section{Model}
\label{secmod}

We consider a linear chain of $N$-qubits with $XX$ interaction. The coupling
strengths are homogeneous
except at one site, the impurity, where the coupling strength is different.
The system is described by the Hamiltonian

\begin{equation}
H(\alpha)=\alpha
J(\sigma_{1}^{x}\sigma_{2}^{x}+\sigma_{1}^{y}\sigma_{2}^{y})+\sum_{i>1}J(\sigma_
{i}^{x}\sigma_{i+1}^{x}+\sigma_{i}^{y}\sigma_{i+1}^{y}),
\label{hamiltonian}
\end{equation}
where $\sigma^{\gamma}$ are the Pauli matrices, $J<0$ is the exchange coupling
coefficient and $\alpha J$
is the impurity exchange strength, $\alpha=1$ corresponds to the
homogeneous case.

Since the Hamiltonian commutes with $S_{z}=\Sigma_{i}\sigma_{i}^{z}$,
the Hamiltonian $H(\alpha)$ has a block structure where each of them
is characterized by the number of excited spins in the chain. Because
we are interested in the transmission of a state with one excited
spin from one end of the chain to the other, we focus on the eigenvectors
of the one excitation subspace where the complete dynamics take place.
To describe the eigenstates, we choose a basis described by the computational
states of this subspace $|n\rangle=(\uparrow\uparrow...\uparrow\downarrow_{n}\uparrow...\uparrow)$,
where $n=1,\ldots,N$ given a basis set size equals to the number
of spins of the chain.

In this basis, the Hamiltonian $H$ is represented by a $N\times N$
matrix 
\begin{equation}
{\mathcal{H}}=\left(\begin{array}{ccccc}
0 & \alpha J & 0 & \dots & 0\\
\alpha J & 0& J & \dots & 0\\
0 & J & 0 & \dots & 0\\
\vdots & \vdots & \vdots & \ddots & J\\
0 & 0 & 0 & J & 0\end{array}\right).\label{hamiltonian matriz}
\end{equation}

Implementations of this model could be realized, for example, with cold atoms
confined in optical lattices
\cite{hartmann2007,dorner2003,lewenstein2007,duan2003} or with nuclear spin
systems in NMR \cite{madi1997,alvarez2010}. While in the first case an initial
pure state in the one excitation subspace can be realized, in the spin ensemble
situations of NMR an effective one excitation subspace is achieved by creating
pseudo pure states where an excess of magnetization is localized on a given
spin.

\section{Energy spectrum and eigenstates}

In this Section we briefly recall some known results about the spectrum
and the eigenstates of the model emphasizing those features that
are of interest in the following Sections.

The one excitation spectrum consists of $N$ eigenenergies denoted
by $\left\{ E_{1}\leq E_{2}\leq...\leq E_{N}\right\} $. Choosing
the total number of spins even the spectrum results symmetrical with respect to
zero  ($E=0$
is not an eigenvalue), for any value of $\alpha$. Then $\left\{
E_{1},...,E_{M}\right\} $
are negative values whereas $\{E_{M+1},...,E_{N}\}$ are positive,
where $M=N/2$. In the homogeneous case ($\alpha=\alpha_{J}\equiv1$),
the energy spectrum lies between the values $\pm2|J|$, this interval is usually
call the {\em band} of eigenvalues. The size
of the chain only changes the number of eigenvalues between those
extreme values, becoming a continuous spectrum when $N\rightarrow\infty$.

\begin{figure}[ht]
\begin{centering}
\includegraphics[width=8cm]{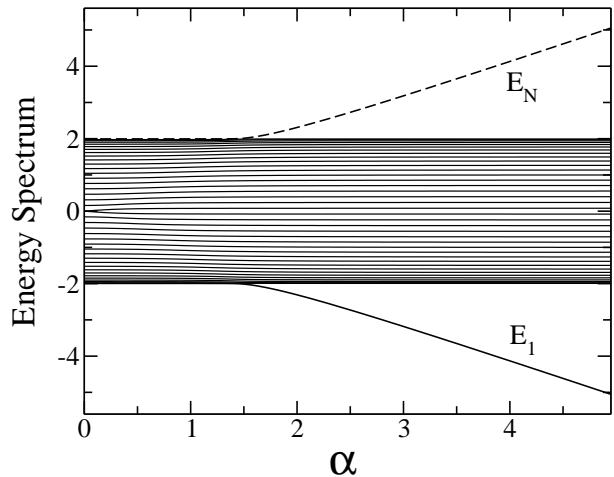}
\par\end{centering}

\caption{\label{energias} The one excitation spectrum {\em vs} $\alpha$
for a spin chain with 40 spins. For $\alpha$ large enough the spectrum
shows two isolated eigenenergies and one band $|E|\leq2|J|$. The
two isolated curves correspond to the minimal eigenenergy $E_{1}$
(continuous line) and the maximal eigenenergy $E_{N}$ (dashed line).
At the critical value $\alpha_{c}$ the isolated energies go into
the band causing a slight distortion on the behaviour of energies
inside the band. In this figure we use $|J|=1$ }
\end{figure}

The inhomogeneous case shows a different behaviour. For large enough
$\alpha$ the minimal and the maximal eigenenergy become isolated
from the band. There is a critical value $\alpha_{c}$ which separates
the region of the spectrum where the energies make a band ($0<\alpha<\alpha_{c}$)
from the region where the energies make a band with two isolated energies
($\alpha>\alpha_{c}$). { The critical point $\alpha_{c}$ can be obtained
analytically, and for large values of $N$,
$\alpha_{c}\simeq\sqrt{2}$. We will further analyze  this point later on.}

For $\alpha\gg\alpha_{c}$  the minimal and maximal energies move
apart from the band proportionally to $-\alpha$ and $+\alpha$ respectively.
This behaviour is depicted in Figure \ref{energias}.

Figure \ref{energias} shows that most of the eigenenergies seem to
be fairly independent of $\alpha$, except for the minimal and maximal
energies. But a more detailed study of the derivative
of the eigenenergies with respect to $\alpha$ (see section V), shows
two regions where the changes in the spectrum are more noticeable:
(i)  for $\alpha\sim0$ two eigenenergies become degenerate because
the system changes from a chain with $N$ coupled spins to a chain
with $N-1$ coupled spins and an uncoupled spin; ii) for
$\alpha\lesssim\alpha_{c}$ there is
a number of avoided crossings between successive eigenenergies, because
of the ``collision'' among the minimal (or maximal) eigenenergy and
the band.

The eigenstates in the one excitation subspace $|\Psi_{E}(\alpha)\rangle$,
whose eigenvalue equation is 
\begin{equation}\label{ecua-autovalores}
H(\alpha)|\Psi_{E}(\alpha)\rangle=E|\Psi_{E}(\alpha)\rangle,
\end{equation}
can be written as a superposition of the one excitation states 
\begin{equation}
|\Psi_{E_{j}}(\alpha)\rangle=\sum_{n=1}^{N}\Psi_{n}^{(j)}\left|n\right\rangle
,\label{one-excitation-sta}
\end{equation}
where due to the symmetries of the spectrum 
\begin{equation}\label{psi-simetricos-1}
{
\Psi_{n}^{(j)}=(-1)^{n}\Psi_{n}^{(N-j+1)}.
}
\end{equation}

These coefficients $\Psi_{n}^{(j)}$ contain information about localization
and entanglement properties of the eigenstates and, can be written
as \cite{santos2003} 
\begin{equation}\label{coe-analiticos}
\Psi_{n}^{(j)}=de^{i\theta n}+d^{\prime}e^{-i\theta n}.
\end{equation}

\begin{figure}[floatfix]
\begin{centering}
\includegraphics[width=8cm]{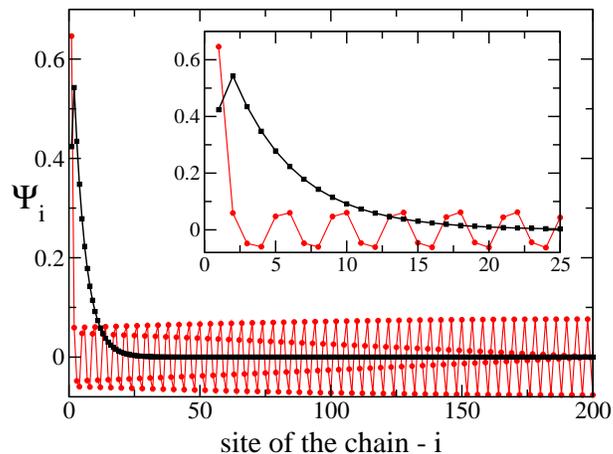}
\par\end{centering}

\caption{\label{two-states}(color online) The coefficients $\Psi_{i}$ for
two different eigenstates, $|\Psi_{E_{1}}(\alpha)\rangle$ with $\alpha=1.6$
(black squares) and $|\Psi_{E_{M}}(\alpha)\rangle$ with $\alpha=0.1$
(red circles). The lines are a guide to the eye. The states and the
values of $\alpha$ were chosen to obtain equal values for their inverse
participation ratios. The inset shows a zoom of the region near $i=1$. }
\end{figure}

In a homogeneous chain, the eigenstates are wave-like superpositions
of the one excitation states  where the coefficients of the superpositions
are given by (\ref{coe-analiticos}) with $\theta$ real. In
other case, $\alpha\ne1$, the eigenstates within the band are very
similar to the states of the homogeneous case (Figure \ref{two-states}
shows $\Psi_{n}^{(M)}$ for $\alpha=0.1$), but they differ in their
coefficient on the impurity site. For $\alpha>\alpha_{c}$ the minimal
eigenenergy state $|\Psi_{E_{1}}\rangle$ is quite different (similarly
for $|\Psi_{E_{N}}\rangle$), its coefficients $\Psi_{n}^{(1)}$ decay
exponentially (Figure \ref{two-states} shows $\Psi_{n}^{(1)}$ for
$\alpha=1.6$).

It is rather simple to show the existence of a localized state when $\alpha\geq
\sqrt{2}$. Using the ansatz $\Psi_1=u_1$ and $\Psi_n=(-1)^{n+1} e^{-n\kappa}$,
for $n\geq 2$, to construct a state $\left|\Psi\right\rangle$, and replacing
this state in Equation~\ref{ecua-autovalores}, after some algebra we obtain
that
\begin{equation}
e^{2\kappa} =  \alpha^2 -1,
\end{equation}
so, to obtain a localized state, the condition $ e^{2\kappa}\geq 1$ implies
that $\alpha\geq \sqrt{2}$. This has been discussed previously see, for
example, the work of Stolze and Vogel \cite{stolze}. In  \cite{stolze} the
authors exploits the mapping between the $XX$ model with one excitation and a
non-interacting fermion model with one particle.

The density matrix for each eigenstate is given by \begin{equation}
\hat{\rho}_{E}(\alpha)=|\Psi_{E}(\alpha)\rangle\langle\Psi_{E}(\alpha)|,\label{rho_{E}}\end{equation}

\noindent which is a $N\times N$ matrix in the one excitation subspace.

\section{Localization of the eigenstates}

As stated above, the eigenenergies and eigenstates change according
to the strength of the impurity considered in the system. To quantify
and study their changes, we calculate the eigenstate localization
as a function of the impurity strength. For that purpose we use the
inverse participation ratio (IPR) \cite{giraud2007}, 
\begin{equation}
L_{IPR}(\left|\Psi\right\rangle )=\frac{\sum_{n}^{^{N}}\Psi_{n}^{2}}{\sum_{n}^{N}\Psi_{n}^{4}},
\end{equation}
where $\Psi_{i}$ are the coefficients of the superposition
(\ref{one-excitation-sta})
of the state. When the state is highly localized ({\em i.e.} $\Psi_{i}$
is nonzero for only one particular value of $i$) $L_{IPR}(\left|\Psi\right\rangle )$
has its minimum value, 1, and when the state is uniformly distributed
(ie. $\Psi_{i}=1/\sqrt{N}$ for all $i$) the IPR attains its  maximum value, $N$.
We call a state $|\Psi\rangle$ {\em extended} if
$L_{IPR}(|\Psi\rangle)\sim{\mathcal{O}}(N)$,
{\em i.e.} the IPR is of the same order
of magnitude than the length of the chain.

\begin{figure}[floatfix]
\begin{centering}
\includegraphics[width=8cm]{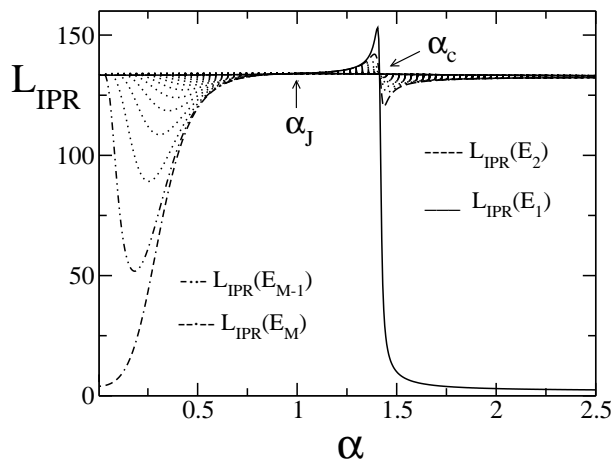}
\par\end{centering}
\caption{\label{IPR1}Localization measure ($L_{IPR}=\sum_{n}\Psi_{n}^{2}/\sum_{n}\Psi_{n}^{4}$)
of different one-excitation eigenstates {\em vs} $\alpha$, for
chain with $N=200$ spins. The values of $\alpha_{J}$ and $\alpha_{c}$
are shown. For $\alpha\gg\alpha_{c}$, the curves of the IPR for the
all eigenstates, except those corresponding to the minimal and maximal
eigenenergies, collapse into a single curve. For $\alpha>\alpha_{c}$
the curves with $L_{IPR}\sim1$ correspond to the minimal and maximal
eigenenergy states. The steep behaviour of these curves when $\alpha\rightarrow\alpha_{c}^{+}$
shows the change from well localized to extended states. The localized
states, with low IPR, that appear for $\alpha<\alpha_{c}$ correspond
to states with eigenenergies near the center of the band. Near $\alpha=0$
there are several localized states. Each curve is double as explained
in the text. }
\end{figure}

From (\ref{psi-simetricos-1}), two states whose eigenenergies
are symmetric with respect to zero, say $|\Psi_{E_{j}}\rangle$ and
$|\Psi_{E_{N-j+1}}\rangle$ where $j\leq N/2$, have the same IPR,
{\em i.e.} $L_{IPR}(|\Psi_{E_{j}}\rangle)=L_{IPR}(|\Psi_{E_{N+1-j}}\rangle)$. As
a consequence, each curve in Figure \ref{IPR1} is double and we consider
the IPR only for the states $\left\{ |\Psi_{E_{1}}\rangle,...,|\Psi_{E_{M}}\rangle\right\} $.

Figure \ref{IPR1} shows the inverse participation ratio $L_{IPR}$
of several eigenstates $\{|\Psi_{E_{1}}\rangle,....,|\Psi_{E_{N}}\rangle\}$
as a function of the impurity coupling $\alpha$. We can identify
three regions where the behaviour of the $L_{IPR}$ is qualitatively
different. These regions are separated by $\alpha_{J}$ and $\alpha_{c}$,
where at those values all eigenstates are equally localized.

The first region $0<\alpha<\alpha_{J}$ shows several localized eigenstates
corresponding to energies close to zero, i.e. the center of the band.
Calling $\alpha_{E_{j}}^{m}$ the value of $\alpha$ such that $L_{IPR}(E_{j},\alpha)=L_{IPR}(\left|\Psi_{E_{j}}(\alpha)\right\rangle )$
attains its minimum, the numerical results show that $L_{IPR}(\alpha_{E_{M}}^{m})<L_{IPR}(\alpha_{E_{M-1}}^{m})<...$
where $\alpha_{E_{M}}^{m}<\alpha_{E_{M-1}}^{m}<...$, {\em i.e.}
the eigenstate is more localized as it is closer to $E=0$. Besides,
the number of localized states increase with $N$.

In the second region $\alpha_{J}<\alpha<\alpha_{c},$ the eigenstates
with energies close to the border of the band become more extended
acquiring a IPR maximum near to $\alpha_{c}$. These peaks become
sharper when $N$ grows. At $\alpha_{c}$, these eigenstates are
again equally localized, but  for values of $\alpha$ larger than
$\alpha_{c}$, but very close to this value, the eigenstates become
more localized.  The size of the interval around $\alpha_{c}$ in
which this critical behaviour can be observed depends on the size of
the chain. This localization changes seem to be related to the avoided
crossings in the spectrum previously described.

In the last region $\alpha>\alpha_{c}$ there are only two eigenstates
highly localized that correspond to the minimal and maximal eigenenergies,
$E_{1}$ and $E_{N}$. The other states are extended through $N-1$
sites of the chain.

We want to stress that the IPR gives a coarse description of the eigenstates,
for example the states in Figure \ref{two-states}, despite of their
very different behaviour, are equally localized if the measure of localization
is the IPR, effectively $L_{IPR}(\Psi_{E_{1}})=L_{IPR}(\Psi_{E_{M}})\simeq5.6$
for both states. This indicates that the IPR can not distinguish
the exponentially localized state from the state with a wave-like
superposition extended over the chain if the latter has its coefficient
$\Psi_{1}$ large enough.

This shows that the IPR is a good tool to quantify changes in the
system due to the introduction of a impurity spin, however it does
not give information about where the eigenstate is localized. Moreover,
it does not distinguish between quite different states as those
described in Figure \ref{two-states}. Studying the coefficients of
the eigenstates, we can observe where they are localized. In the
present case they are mainly localized on the impurity site (see
Figure \ref{two-states}). However, since we are interested in the transmission of
initially localized quantum states, and that a successful transmission results
in another localized state, the IPR could provide an easy way to identify when
the transmission has taken place.

Since the IPR does not distinguish between the exponentially localized
states that lie outside the band of magnons and the localized states inside the
band it is necessary to study both kinds of states using a local quantity. In
the next Section we study the entanglement between the impurity site and its
first neighbor, this will allow us to classify the different eigenstates
accordingly with its entanglement content.

\section{Entanglement of the eigenstates}

The bipartite entanglement between two qubits can be calculated using
the Concurrence \cite{woottersPRL98}. The Concurrence of two qubits
in an arbitrary state characterized by the density matrix $\rho$
is given by

\begin{equation}
C(\rho)=max\{0,\lambda_{1}-\lambda_{2}-\lambda_{3}-\lambda_{4}\},\label{concu}\end{equation}
 where the $\lambda_{i}$ are the square roots of the eigenvalues,
in decreasing order, of the non-Hermitian matrix $\rho\widetilde{\rho}$.
The spin-flipped state $\widetilde{\rho}$ is defined as
\begin{equation}
\widetilde{\rho}=(\sigma^{y}\otimes\sigma^{y})\rho^{*}(\sigma^{y}\otimes\sigma^{y}),
\end{equation}
were $\rho^{*}$ is the complex conjugate of $\rho$ and it is taken
in the computational basis $\{|\uparrow\uparrow\rangle,
|\uparrow\downarrow\rangle\, |\downarrow\uparrow\rangle, 
|\downarrow\downarrow\rangle\}$.
The concurrence takes values between 0 and 1, where 0 means that the
state is disentangled whereas 1 means a maximally entangled state.

\noindent When considering a subsystem of two qubits on the chain,
the concurrence is calculated with the reduced density matrix. The
reduced density matrix for the spin pair $(i,j)$, $\rho_{E}^{(i,j)}(\alpha)$,
corresponding to the eigenstate $|\Psi_{E}(\alpha)\rangle$ is given
by

\begin{equation}
\rho_{E}^{(i,j)}(\alpha)={Tr}\left|\Psi_{E}(\alpha)\right\rangle \left\langle \Psi_{E}(\alpha)\right|={\mathrm{T}r}\hat{\rho}_{E}(\alpha),
\end{equation}

\noindent where the trace is taken over the remaining $N-2$ spins leading
to a $4\times4$ matrix.

The structure of the reduced density matrix follows from the symmetry
properties of the Hamiltonian. Thus, in our case the concurrence
$C(\rho_{E_{k}}^{(i,j)})$ depends on $i$ and $j$, {\em i.e.}
the indexes of the sites where the spin pair lies. Note that in the
translationally invariant case $C(\rho_{E_{k}}^{(i,j)})$ depends
only on $|i-j|$. In what follows
$C_{i,j}=C_{i,j}(\rho_{E_{k}})=C(\rho_{E_{k}}^{(i,j)})$.

Using the definition $\langle\hat{A}\rangle=Tr(\hat{\rho}\hat{A})$,
we can express all the matrix elements of the density matrix $\rho^{(i,j)}$
in terms of different spin-spin correlation functions. In particular,
for nearest neighbors spins and the eigenstate $|\Psi_{E_{j}}\rangle$,
we get

\begin{equation}
\rho_{E_{j}}^{(i,i+1)}=\left(\begin{array}{cccc}
a_{j} & 0 & 0 & 0\\
0 & b_{j} & \langle\sigma_{i}^{+}\sigma_{i+1}^{-}\rangle_{E_{j}} & 0\\
0 & \langle\sigma_{i}^{+}\sigma_{i+1}^{-}\rangle_{E_{j}}^{*} & d_{j} & 0\\
0 & 0 & 0 & 0\end{array}\right),\label{mat-red}\end{equation}
 where \begin{equation}
a_{j}=\frac{1}{4}\langle(\sigma^{z}+I)_{i}(\sigma^{z}+I)_{i+1}\rangle_{E_{j}},\end{equation}

\begin{equation}
b_{j}=\frac{1}{4}\langle(\sigma^{z}+I)_{i}(I-\sigma^{z})_{i+1}\rangle_{E_{j}},\end{equation}

\begin{equation}
d_{j}=\frac{1}{4}\langle(I-\sigma^{z})_{i}(\sigma^{z}+I)_{i+1}\rangle_{E_{j}},\end{equation}
 $I$ is the $2\times2$ identity matrix, $\sigma_{i}^{\pm}=(\sigma_{i}^{x}\pm i\sigma_{i}^{y})/2$,
and 
\begin{equation}
\langle\ldots\rangle_{E_{j}}=\left\langle \Psi_{E_{j}}\right|\ldots\left|\Psi_{E_{j}}\right\rangle.
\end{equation}
Thus, the concurrence results to be \begin{equation}
C_{i,i+1}(\rho_{E_{j}})=\max\{0,2\mid\langle\sigma_{i}^{+}\sigma_{i+1}^{-}\rangle_{E_{j}}\mid,2\sqrt{|b_{j}d_{j}|}\}.\label{formula_concurrence}
\end{equation}
 For the set of eigenstates that we are considering, the expression
for the concurrence can be further simplified. After some algebra we get
\begin{equation}
b_{j}=(\Psi_{i+1}^{(j)})^{2},\quad d_{j}=(\Psi_{i}^{(j)})^{2},
\end{equation}
and that 
\begin{equation}
\langle\sigma_{i}^{+}\sigma_{i+1}^{-}\rangle_{E_{j}}=\Psi_{i+1}^{(j)}\Psi_{i}^{(j)}.
\end{equation}

So, we get that
\begin{equation}
C_{i,i+1}(\rho_{E_{j}})=2\left|\Psi_{i+1}^{(j)}\Psi_{i}^{(j)}\right|.
\end{equation}

Using the Hellmann-Feynman theorem, and the symmetry properties of
the Hamiltonian, we find that 
\begin{equation}
\frac{\partial E_{j}}{\partial\alpha}=2J\left\langle \Psi_{E_{j}}\right|\sigma_{1}^{+}\sigma_{2}^{-}\left|\Psi_{E_{j}}\right\rangle.\label{no-diagonal}
\end{equation}
 From the expression for the reduced density matrix $\rho^{(i,i+1)}$,
(\ref{mat-red}), it is clear that when $\langle\sigma_{i}^{+}\sigma_{i+1}^{-}\rangle=0$
the reduced density matrix is diagonal and the bipartite entanglement
is zero. Moreover, from (\ref{no-diagonal}), when $\frac{\partial E_{j}}{\partial\alpha}=0$
we have that $C_{12}(\rho_{E_{j}})=0$.

So, the concurrence for the first two spins in the eigenstate $|\Psi_{E_{j}}\rangle$
is given by 
\begin{equation}
C_{12}=\left|\frac{1}{J}\frac{\partial E_{j}}{\partial\alpha}\right|.\label{c12-energia-coe}
\end{equation}

We are interested in the relationship between localization and entanglement
for the whole one spin excitation spectrum. In particular, we want to show that
the bipartite entanglement of a given eigenstate, which is a local quantity,
between the impurity site and its first neighbor detects the type of
localization that the eigenstate possess.

\begin{figure}[floatfix]
\begin{centering}
\includegraphics[width=7.5cm]{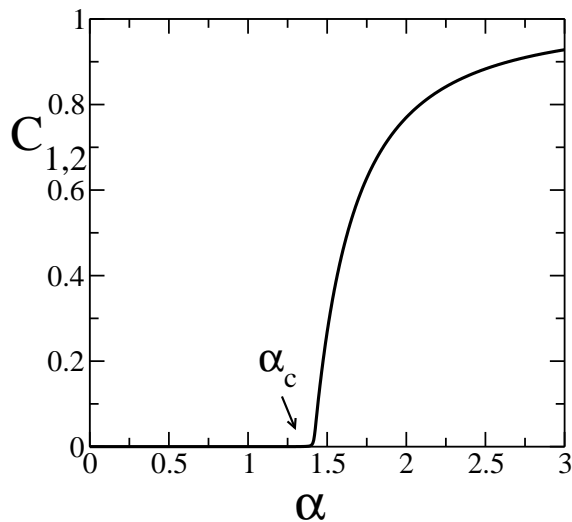}
\par\end{centering}
\caption{\label{concurrence01} Entanglement between the first spin (the impurity
site) and its nearest neighbor for the eigenstate of the minimal
eigenenergy $E_{1}$. It is measured by the concurrence $C_{1,2}(\rho_{E_{1}})$
as a function of $\alpha$. When the state is localized, $\alpha>\alpha_{c}$,
spins 1 and 2 are also entangled. Before the critical point($\alpha\leq\alpha_{c}$)
when the state is extended, $C_{1,2}(\rho_{E_{1}})=0$ consistently
with $\frac{\partial E_{1}}{\partial\alpha}=0$ for $\alpha\leq\alpha_{c}$
. }
\end{figure}

First, we proceed to analyze the concurrence of the minimal eigenenergy state,
$C_{1,2}(\rho_{E_{1}})$
as a function of $\alpha$, the behaviour of this quantity is shown in
Figure \ref{concurrence01}. At first sight, it is clear that
$C_{1,2}(\rho_{E_{1}})$ is different from zero where
$L_{IPR}(\left|\Psi_{E_{1}}\right\rangle )$ (see Figure \ref{IPR1}) is
noticeable, and that $C_{1,2}(\rho_{E_{1}})\rightarrow 0$ when the eigenvalue
enters into the band and, consequently, the eigenstate becomes extended.

So, when the minimal eigenenergy state is
extended for $\alpha<\alpha_{c}$, the two first spins are disentangled
and $C_{1,2}(\rho_{E_{1}})=0$ consistently with $\frac{\partial E_{1}}{\partial\alpha}=0$
from (\ref{c12-energia-coe}). At the critical point $\alpha_{c}$,
the state starts to become localized increasing its degree of localization
when $\alpha\gg\alpha_{c}$; in the same way, the pair of spins starts
to became entangled and almost disentangled from the rest of the chain,
i.e. $C_{1,2}(\rho_{E_{1}})\sim1$.

Actually, the data shown in Figure \ref{concurrence01} corresponds too to
$C_{1,2}(\rho_{E_{N}}(\alpha))$, this can be seen 
by the following argument.

As in the case of the IPR, the concurrence $C_{12}$ for eigenstates
with symmetrical eigenenergies respect to zero ($E_{j}$ and $E_{N-j+1}$)
is the same. From Eqs.~(\ref{psi-simetricos-1}) and
(\ref{c12-energia-coe}),
it is straightforward to demonstrate the latter affirmation where 
\begin{equation}
C_{12}(\rho_{E_{j}})=C_{12}(\rho_{E_{N-j+1}}),\quad j=1,\ldots,M,
\end{equation}
 since
 \begin{equation}
\frac{\partial E_{j}}{\partial\alpha}=-\frac{\partial E_{N-j+1}}{\partial\alpha}.\end{equation}

\begin{figure}[floatfix]
\begin{centering}
\includegraphics[width=8cm]{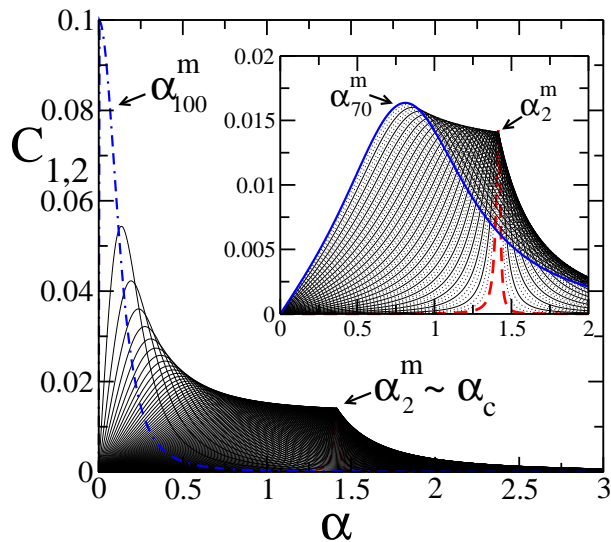}
\par\end{centering}

\caption{\label{concurrence03}(color online)  Concurrence $C_{12}(\rho_{E_{j}}(\alpha))$
as a function of the impurity strength $\alpha$, for $j=2,3,\ldots,100$.
The results were obtained for a chain of $N=200$ spins. Each curve
$C_{12}(\rho_{E_{j}}(\alpha))$ has a single peak. The peaks are ordered
by eigenenergy, the rightmost peak corresponds to $C_{12}(\rho_{E_{2}}(\alpha))$
(red dashed line), the peak to its left corresponds to $C_{12}(\rho_{E_{3}}(\alpha))$,
and so on. The leftmost peak corresponds to the curve with the highest
eigenenergy shown in the figure, $E_{100}$ (blue dashed-dotted line),
belong to the energy of the center of the band. The inset shows the concurrence
$C_{12}(\rho_{E_{j}}(\alpha))$ for $j=2,\ldots 70$}

\end{figure}

Following with the analysis of the entanglement between the first
two spin in the chain, we calculate the concurrence of the states
with energies inside the bands. Figure~\ref{concurrence03} shows
$C_{12}(\rho_{E_{j}})$ as a function of $\alpha$ for $j=2,\ldots,M$.
Note that the same scenario is observed for $C_{12}(\rho_{E_{j}})$
with $j=N-1,\ldots,M+1$.

From Figure \ref{concurrence03}, and calling $\alpha_{i}^{m}$
the abscissa where $C_{12}(\rho_{E_{i}}(\alpha))$ has its maximum, we observe
that
$\alpha_{M}^{m}<...<\alpha_{2}^{m}$ and
$C_{12}(\rho_{E_{M}}(\alpha_{M}^{m}))>....>C_{12}(\rho_{E_{2}}(\alpha_{2}^{m}
))$.
This observation suggests that the ordering of the maxima of the concurrence
$C_{12}$
for the different eigenstates follows closely the ordering dictated by the 
amount of localization of these eigenstates, {\em i.e} only the most localized 
states around the impurity site has a noticeable entanglement. We will use this 
observation as a guide to formulate a transmission protocol in the next Section.

As we have shown, the concurrence and the derivative of the energy are related
in a simple way, see (\ref{c12-energia-coe}). On the other hand it is well
known that the eigenvalues $E_i(\alpha)$ inside the band are rather insensitive
to changes in $\alpha$, indeed  $\partial E_i(\alpha)/\partial \alpha \simeq 0$
almost everywhere, {\em except} near an avoided crossing with other eigenvalue.
In this sense, the behaviour shown by the concurrence in
Figure~\ref{concurrence03} reflects the presence of successive avoided
crossings between $E_1(\alpha)$ and $E_2(\alpha)$, between $E_2(\alpha)$ and
$E_3(\alpha)$, and so on. The abscissa of the peak in the concurrence of a
given eigenstate roughly corresponds to the point where the eigenvalue becomes
almost degenerate. 

As a matter of fact, the scenario depicted in Figure~\ref{concurrence03} is not
only a manifestation of the avoided crossings in the spectrum, indeed it can be
considered as a precursor of the resonance state that appears in the system
when $N\rightarrow \infty$. Recently, Ferr\'on {\em et al.}~\cite{ferron2009}
have shown how the behaviour of an entanglement measure can be used to
detect a resonance state. In a chain a resonance state appears in the limit
$N\rightarrow\infty$, however the peaks in the concurrence obtained for 
$N$ large, but finite, can be used to obtain
approximately the energy of the resonance state~\cite{ferron2009,pont2010}.

\section{Transmission of states and entanglement}

\begin{figure}[floatfix]
\begin{centering}
\includegraphics[width=11cm]{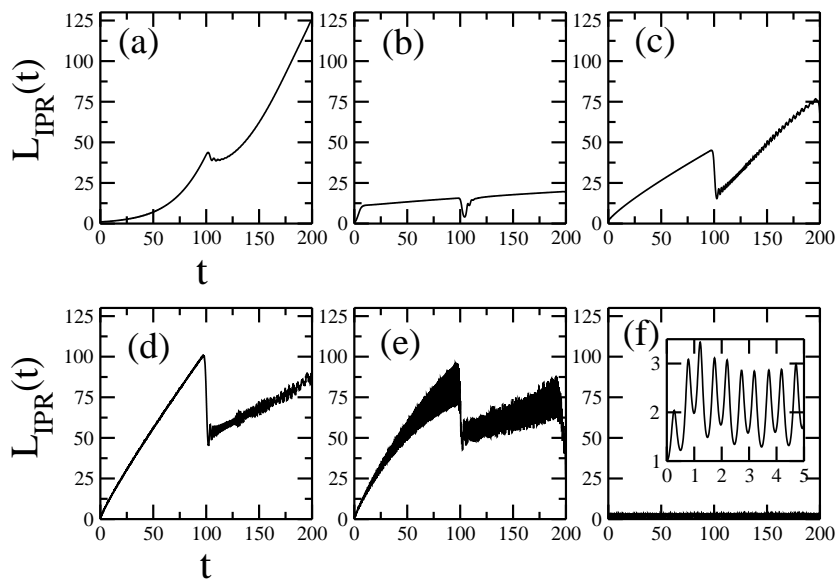}
\par\end{centering}

\caption{\label{ipr-vs-t}The panels show the dynamical behaviour of $L_{IPR}$
{\em vs} $t$, for different values of $\alpha$. a) $\alpha=0.1$,
b) $\alpha=0.4$, c) $\alpha=1$, d) $\alpha=1.4$, e) $\alpha=1.5$
, and f) $\alpha=3$. In all the cases $|1\rangle$ is the initial
condition. The inset in f) shows the small oscillations that characterize
the behaviour of $L_{IPR}$ for $\alpha=3$, in this case the state
of the system is localized even for very long times. In f) the initial
excitation goes back and forth between the impurity site and the
rest of the chain with a frequency given, basically, by the difference
of energy between the two lowest eigenenergies. The steep change near
$t\sim100$, that can be observed in all the panels except in f),
signals the {}``arrival'' of the excitation to the end of the
chain. Note that the refocusing, {\em i.e.} that the value of $L_{IPR}$
drops, is different in each regime, but in b) the refocusing leads
to $L_{IPR}\sim O(1)$. The results were obtained for a chain with
$N=200$.}
\end{figure}

The effect of the localized states in the one magnon band is best
appreciated looking at the dynamical behaviour of the inverse participation
ratio. Figure~\ref{ipr-vs-t} shows the behaviour of $L_{IPR}(|\psi(t)\rangle)$,
where $|\psi(t)\rangle$ satisfies that \begin{equation}
i\frac{d|\psi(t)\rangle}{dt}=H|\psi(t)\rangle,\quad|\psi(t=0)\rangle=|1\rangle,\end{equation}
for different values of $\alpha$. There are, at least, three well
defined dynamical behaviours, each one associated to the number of
localized states in the system, see Figure~\ref{IPR1}. Figure~\ref{ipr-vs-t}
a) ($\alpha=0.1$) shows the behaviour of $L_{IPR}$ when there is
only one localized state at the center of the band; Figure~\ref{ipr-vs-t}
b) ($\alpha=0.4$) shows the dynamical behaviour of $L_{IPR}$ when
there are several localized states; the panels c), d) and e) show
the dynamical behaviour near the transition zone and, finally, f)
shows the dynamical behaviour when the system have exponentially localized
states.

We do not want to analyze completely the rich dynamical behaviour
of $L_{IPR}$, however, from the point of view of the transmission
of quantum states, it is clear that the regime shown in panel b) seems
to be particularly useful. The panel b) shows that when the system
has several localized eigenstates $|\psi(t)\rangle$ consists in
a superposition of a reduced number of elements of the one excitation
states, {\em i.e} the number of significant coefficients $\Psi_{i}$
is small compared with $N$. Besides, the refocusing of the state
when the {}``signal'' reaches the end of the chain (near $t\simeq100$)
leads to an smaller $L_{IPR}$ when $\alpha=0.4$ than for the other
values of $\alpha$, compare panel b) with a), c), d) and e). The
case shown in f) is rather different, in this case the superposition
between the initial state $|1\rangle$ and the localized state is
rather big, so $|\psi(t)\rangle$ remains localized even for very
long times. This dynamical regime has been proposed to store quantum
states \cite{apollaro2006} and, more generally, this kind of states
with isolated eigenvalues has been proposed as a possible scenario
to implement practically a stable qubit \cite{michoel2009}.

We want to remark some points: 1) for very small $\alpha$ there is a
{}``refocusing'' such that $L_{IPR}\sim1$ for $t\sim{\mathcal{O}}(10^{4})$
when $N=200$. 2) The initial excitation that is localized in the
impurities diffuses over the chain \cite{Dente2008} so, for a given time $t$, the
number of sites on the chain that are excited is given, approximately,
by $L_{IPR}(t)$. The presence of localized states reduces this
number and the speed of propagation. For $0.3\lesssim\alpha<\alpha_{c}$
the refocusing of the signal appears a $t\sim N/2$, this time is
roughly independent of $\alpha$. For $\alpha\lesssim0.3$ the time
behaviour is more complicated but the refocusing times scales as $1/\alpha$,
approximately, for fixed $N$, we will consider back this last point
later.

We will use the regime b) identified in Figure~\ref{ipr-vs-t} to
implement the simplest transmission protocol, as suggested by Bose
\cite{bose2008,bose2003}, and the transmission of an entangled state.
But, as our results suggest, we will place a second impurity at the
end of the chain where the transmission should be detected. Locating
an impurity at the end of the chain introduces a set of localized
states around this site. The overall properties of the spectrum do
not change, however the presence of localized states at the end of
the chain would facilitate the transmission of states (or entanglement)
from one end of the chain to the other.

\begin{figure}[floatfix]
\begin{centering}
\includegraphics[width=8cm]{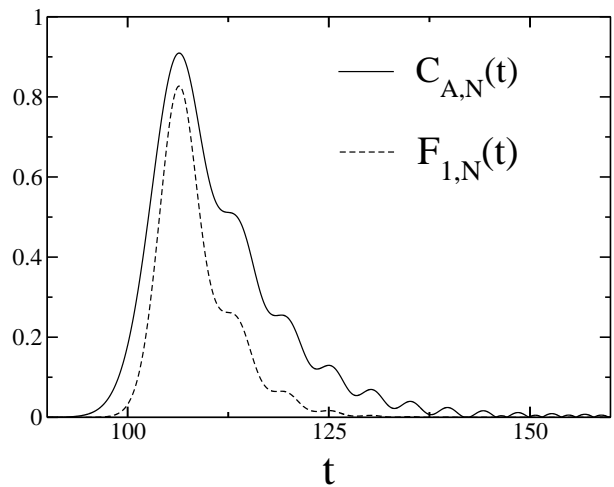} 
\par\end{centering}

\caption{\label{con-and-fidel} The concurrence, $C_{A,N}$ (solid black line)
and the fidelity $F_{1,N}$ (dashed black line) {\em vs} $t$ for
a chain with $N=200$ spins. }
\end{figure}

In the simplest protocol of transmission (as described in \cite{bose2003})
the initial state, $|1\rangle$ evolves following the Hamiltonian dynamics,
and the quality of the transmission is measured with the fidelity
\begin{equation}
{F}=\langle1|\rho_{out}(t)|1\rangle,
\end{equation}
where $\rho_{out}(t)$ is the state at the end of the chain where
the transmission is received, and $t$ is the {}``arrival'' time.

For the transmission of an entangled state the protocol is slightly
different, again we follow the protocol described in \cite{bose2003}.
Using an auxiliary qubit $A$, and the first spin of the chain, the
state \begin{equation}
|\psi^{+}\rangle=\frac{1}{\sqrt{2}}(|\uparrow_{A}\downarrow_{1}\rangle+|\downarrow_{A}\uparrow_{1}\rangle)\end{equation}
is prepared. After the preparation of the initial state the systems
evolves accordingly with its Hamiltonian and the concurrence between
$A$ and the spin at the receiving end of the chain, $C_{A,N}(t)$,
is evaluated.

\begin{figure}[floatfix]
\begin{centering}
\includegraphics[width=12cm]{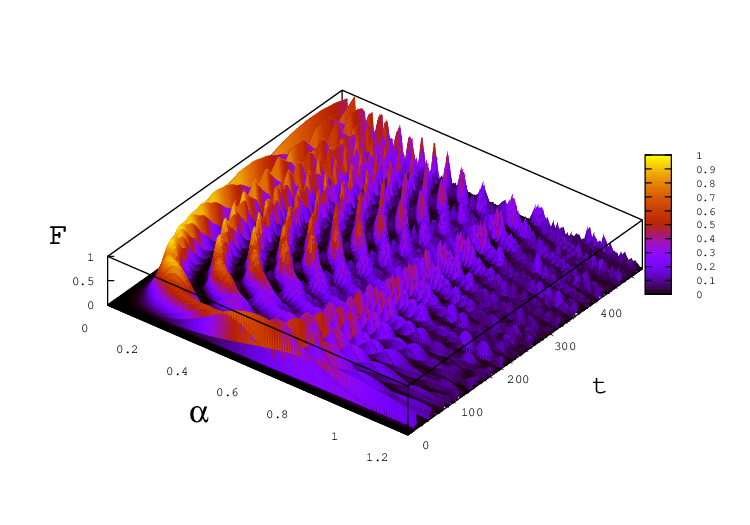} 
\par\end{centering}

\caption{\label{landscape} The fidelity of transmission versus the strength
of the impurities and time. The fidelity presents a peak near $\alpha\simeq0.6$ and $t\simeq15$ for $N=31$. In this peak the
fidelity is rather big.}

\end{figure}

Figure~\ref{con-and-fidel} shows the fidelity for the simplest transmission
protocol and the concurrence between the auxiliary qubit and the last
spin of the chain both as functions of the time. The strength of
the interaction between the first and the second spin is the same
that between the last and its neighbor, $\alpha J$, with $\alpha=0.4$,
and the chain has $N=200$ spins. The maximum value of the fidelity
and the concurrence are remarkably high. For our chain $C_{max}\simeq0.9$,
while for an unmodulated chain (with 200 spins) $C_{max}^{un}\simeq0.23$
\cite{bose2008,bose2003}. It is worth to remark that this large
value of the fidelity is not necessarily the larger possible tuning
the value of $\alpha$.

As a matter of fact, that a chain with two symmetrical impurities
outperforms a homogeneous one as a transmission device has been already
reported in \cite{wojcik2005}. In that work, W\'ojcik {\em
et al.} analyzed the transmission of quantum states modulating the
coupling between the source and destination qubits. They shown that
using small values of the coupling it is possible to obtain a fidelity
of transmission arbitrarily close to one with the transfer time scaling
linearly with the length $N$. Regrettably the resulting transfer
time obtained in their work is quite large. Here we will extend their
results showing that the transfer of quantum states is feasible for
shorter transfer times with a very good fidelity {( $\gtrsim$ 0.9)}
while keeping the linear scaling between the transfer time and the
length of the chain. To achieve this transfer scenario we will exploit
the information provided by the IPR: for large enough values of $\alpha$
there is a time of order $N/2$ such that $L_{IPR}\sim1$.

\begin{figure}[floatfix]
\begin{centering}
\includegraphics[width=11cm]{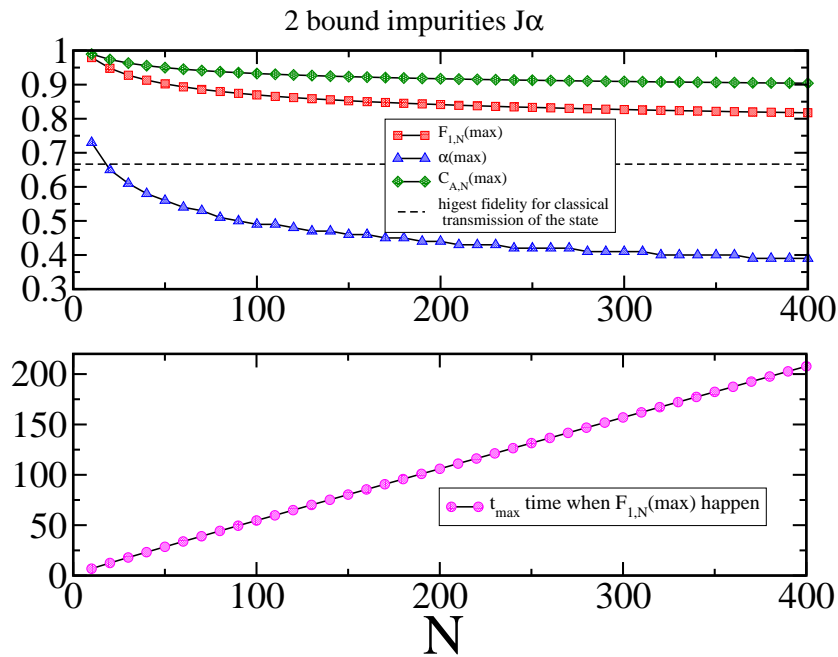} 
\par\end{centering}

\caption{\label{transfer-all} The data shown in the upper panel corresponds
to the maximum fidelity of transmission achievable for times $t_{tr}\sim t_{IPR}\sim{\mathcal{O}}(N/2)$
for different chain lengths $N$ (squares), $\alpha_{opt}(N)$ (triangles)
and the concurrence $C_{A,N}$(diamonds). The protocol of transmission
is described in the text. }
\end{figure}

The identification of regimes where the transmission of quantum states
can be achieved with large fidelity and for (relatively) short times
is of great importance. The different dynamical regimes of the fidelity
in a chain with two impurities is rather difficult to analyze except
when $\alpha\rightarrow0$, see \cite{wojcik2005}. Figure~\ref{landscape}
shows the complex landscape of the fidelity of transmission versus
the strength of the impurities and time. Some of the features shown
by the fidelity in Figure~\ref{landscape} are best understood using
the IPR. In particular, for $\alpha$ fixed, the first maximum of
the fidelity as a function of the time coincides with a minimum of
$L_{IPR}$. This observation, once systematized, provides the dynamical
regime where the transmission can be achieved with large fidelity
and {\em always} for times $\sim N/2$.

Our results about the time behaviour of the IPR show that for
$t_{IPR}\sim{{\mathcal{O}}}(N/2)$
there is always a local minimum of the IPR (see Figure~\ref{ipr-vs-t}
b)). Since the state that is being transferred is well localized it
is rather clear that we should look for times when the IPR attain
local minima to identify where it is possible to achieve a good transmission.
The time $t_{IPR}$ is rather independent of $\alpha$. So, optimizing
the value of $\alpha$ in order to minimize the value of the minimum
of the IPR at times $\sim t_{IPR}$ allow us to find the best fidelity
achievable for time $t_{tr}\sim t_{IPR}$. We call $\alpha_{opt}(N)$
the value such that the the fidelity $F(t_{tr})$ attains its maximum
for a given $N$ and for $t_{tr}\sim N/2$.

As Figure~\ref{con-and-fidel} shows, when the transfer of a given
state takes place the fidelity presents a well defined maximum at
time $t_{tr}\sim t_{IPR}\sim{\mathcal{O}}(N/2)$. The height of the
maximum, $F_{max}$ is a smooth function of $\alpha$ for $\alpha>0.3$,
and the same is valid for the transfer time $t_{tr}$.

Figure~\ref{transfer-all} summarizes our findings about the fidelity
of transmission following the recipe outlined in the two paragraphs
above. The upper panel shows the maximum transmission fidelity achievable
for a chain of length $N$ and the corresponding optimum value of
$\alpha$. As can be appreciated $F\gtrsim0.8$ even for $N=400$.
The maximum value of the fidelity is also well above the predicted
for an unmodulated chain and above $2/3$ that is the highest fidelity
for classical transmission of a quantum state. The lower panel shows
the transmission time $t_{tr}$ {\em vs} $N$. The linear scaling
of $t_{tr}$ with $N$ is rather clear.

It is clear that for an isolated chain the availability of a regime
where $F\sim1$, regardless of the time required to achieve the transfer,
is interesting. However, in the presence of dynamical disorder or
an {}``environment'', achieving a moderate fidelity for the transfer
at shorter times seems a better option.

\section{Discussion}

\label{discu}

There is enough evidence to affirm that the entanglement of quantum
states whose eigenenergies present avoided crossings will show steep
changes near of them (\cite{ferron2009},\cite{sha-inf-avoi}, this
work). In our case there is a number of avoided crossings that appears
between successive levels, when $E_{1}$ comes into the band as $\alpha$
decreases from values larger than the critical. The avoided crossing
between $E_{1}$ and $E_{2}$ is nearer to $\alpha_{c}$ than the
avoided crossing between $E_{2}$ and $E_{3}$, and so on. This is
the behaviour depicted in Figure (\ref{concurrence03}). The width
of the peak in $C_{12}$ of a given state (see Figure (\ref{concurrence03}))
is related to the magnitude of the derivative of the eigenenergy
of the state, the peak is sharper for $C_{12}(\rho_{E_{2}})$ and
the successive peaks are more and more rounded.

As we have shown, locating impurities at both extremes of the chain
allows to transfer more entanglement that an unmodulated chain {\em
if both impurities produce a number of localized eigenstates at each
end of the chain}. If a initially localized state is transmitted through the
chain, at a posterior time  the state is composed by the superposition of many
propagating modes. The optimization of the couplings at the end of the chain
allows
the coherent superposition of many of those modes at some time
$t_{tr}$, resulting in a large fidelity of transmission. The arrival time
$t_{tr}$ is  {\em always} $\geq N/2$. It could be interesting to compare the
results
presented in this work with the findings of Plastina and Apollaro
(\cite{plastina2007}) in the case of two {\em diagonal} impurities.

While IPR is an appealing quantity since it is very easy to calculate,
we have shown that it is not possible to guess how much entanglement
has a given state. The examples analyzed show that based on the IPR
it is not possible to guess from it how much entanglement has a given
state, anyway it remains an appealing quantity since it could be useful
to identify dynamical regimes where the transmission of quantum states
can be achieved. The example presented above, in which the tuning
of the interaction between only a couple of spins improves the transmission,
is encouraging. Of course the protocols for perfect transmission
perform this task better, but at the cost of modulating all the interactions
between the spins.

There is not, to our knowledge, a simple quantity that allows to relate,
in a direct way, localization and entanglement. This subject will
be object of further investigation.

\ack We would like to acknowledge SECYT-UNC 05/B337,
CONICET PIP 112-200801-01741, and FONCyT Grant N$^{o}$ PICT-2005
33305 (Argentina) for partial financial support of this work. We thank
B. Franzoni and G. A. Alvarez for helpful discussions and critical reading of the
manuscript.We thank Dr T Apollaro that very recently brought our attention to 
reference \cite{Banchi2010} which deals with the subject addressed in this work.

\end{document}